\documentclass[fleqn,twoside]{article}
\usepackage{espcrc2,epsf,epsfig,psfrag,graphicx}
\usepackage[figuresright]{rotating}
\newcommand{\be}{\begin{equation}}
\newcommand{\ee}{\end{equation}}
\newcommand{\bea}{\begin{eqnarray}}
\newcommand{\eea}{\end{eqnarray}}
\def\aprle{\buildrel < \over {_{\sim}}}
\def\aprge{\buildrel > \over {_{\sim}}}

\newcommand{\AmS}{{\protect\the\textfont2
  A\kern-.1667em\lower.5ex\hbox{M}\kern-.125emS}}

\title{Neutrino oscillations beyond two flavours }

\author{E. Kh. Akhmedov\address[IST]{Centro de F\'\i sica das Interac\c
c\~oes Fundamentais 
\\ 
Departamento de F\'\i sica, Instituto Superior T\'ecnico \\
Av. Rovisco Pais, P-1049-001 Lisboa, Portugal}
        \thanks{On leave from National Research Centre Kurchatov
Institute, Moscow, Russia. 
Supported by the Calouste Gulbenkian Foundation as a 
Gulbenkian Visiting Professor. 
}}       

\begin{document}

\begin{abstract}
I review some theoretical aspects of neutrino oscillations in the
case when more than two neutrino flavours are involved. These include: 
approximate analytic solutions for 3-flavour (3f) oscillations in matter; 
matter effects in $\nu_\mu \leftrightarrow \nu_\tau$ oscillations; 3f 
effects in oscillations of solar, atmospheric, reactor and supernova 
neutrinos and in accelerator long-baseline experiments; CP and T violation 
in neutrino oscillations in vacuum and in matter; the problem of $U_{e3}$; 
4f oscillations.  

\vspace{1pc}
\end{abstract}

\maketitle

\section{INTRODUCTION}

Explanation of the solar and atmospheric neutrino data in terms of
neutrino oscillations requires at least three neutrino species, and 
in fact three neutrino species are known to exist -- $\nu_e$, $\nu_\mu$ 
and $\nu_\tau$. 
%
%
If the LSND experiment is correct, then probably a fourth neutrino type 
should exist, a light sterile neutrino $\nu_s$. However, until
relatively recently most of the studies of neutrino oscillations 
were performed in the 2-flavour framework. There were essentially two 
reasons for that: (1) simplicity -- there are much fewer parameters in the
2-flavour case than in the 3-flavour one, and  the expressions for the
transition probabilities are much simpler and by far more tractable, 
and (2) the hierarchy of $\Delta m^2$ values, which allows to effectively
decouple different oscillation channels. The 2-flavour approach proved to
be a good first approximation, 
which is a consequence of the hierarchy $\Delta m_\odot \ll 
\Delta m_{\rm atm}$ and of the smallness of the leptonic mixing parameter 
$|U_{e3}|$. 

However, the increased accuracy of the available and especially forthcoming 
neutrino data makes it very important to take into account even relatively 
small effects in neutrino oscillations. In addition, the experimentally 
favoured solution of the solar neutrino problem is at present the LMA MSW 
one, which requires the hierarchy between $\Delta m_\odot$ and $\Delta 
m_{\rm atm}$ to be relatively mild. Also, effects specific to $\ge 3$ 
flavour neutrino oscillations, such as CP and T violation, are now being 
very widely discussed. All this makes 3-flavour (or 4-flavour) analyses of 
neutrino oscillations mandatory. 

In my talk I review some theoretical issues pertaining to neutrino
oscillations in the case when more than two neutrino species are involved. 
I mainly concentrate on 3-flavour (3f) oscillations and only very briefly 
consider the 4f case. The topics that are discussed include: approximate 
analytic solutions for 3f oscillations in matter; matter effects in 
$\nu_\mu \leftrightarrow \nu_\tau$ oscillations; 3f effects in oscillations
of solar, atmospheric, reactor and supernova neutrinos and in accelerator 
long-baseline experiments; CP and T violation in neutrino oscillations in 
vacuum and in matter;    
the problem of $U_{e3}$; 4f oscillations.  

\section{\mbox{
NEUTRINO 
OSCILLATIONS IN}\\ \hspace*{1.5mm}MATTER (3f)}
%
%
Neutrino oscillations in matter are described by the
evolution equation $i(d/dt)\nu = H \nu$, 
where $\nu=(\nu_e ~\nu_\mu ~\nu_\tau)^T$ and 
\be
H\!=\!\!\left[U \! \left( \begin{array}{ccc} \!\! E_1 \! & \! 0 \! 
& \! 0 \! \\ \!\! 0 \! & \! E_2 \! & \! 0 \! \\ \!\! 0 \! & \! 0 
\! & \! E_3 \! \end{array} 
\right) \! 
U^\dagger
+ \left( \begin{array}{ccc}\!\! V(t) \!\! & \! 0 & 0 \! \\ \!\! 0 \!\! 
& \! 0 & 0 \! \\ \!\! 0 \!\! & \! 0 & 0 \!
\end{array} \right) \! \right]\!\!
\label{Sc1}
\ee
The effective potential $V=\sqrt{2}G_F N_e$ is due to the charged-current 
interaction of $\nu_e$ with the electrons of the medium. The neutral current 
induced potentials are omitted from Eq. (\ref{Sc1}) because they are the same 
for neutrinos of all three species and therefore do not affect neutrino 
oscillations. This, however, is only true in leading (tree) order; radiative 
corrections induce tiny differences between the neutral current potentials 
of $\nu_e$, $\nu_\mu$ and $\nu_\tau$ and, 
in particular, result in a very small $\nu_\mu$ -- $\nu_\tau$ 
potential difference $V_{\mu\tau}\sim 10^{-5}\, V$ \cite{BLM}. 
This quantity is negligible in most situations but may be important for 
supernova neutrinos.

\begin{figure}[t] 
    \includegraphics[width=0.46\textwidth]{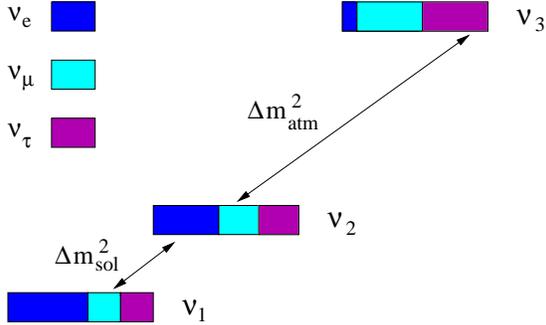}
\vspace*{-4mm}
    \caption{\label{fig:schemes}Normal mass hierarchy }   
\label{scheme1}
\vspace*{-2mm}
\end{figure}
%
\begin{figure}[t] 
    \includegraphics[width=0.46\textwidth]{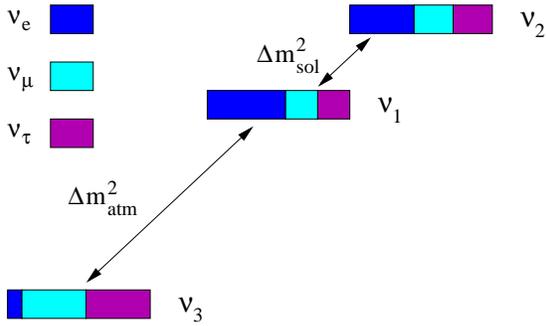}
\vspace*{-4mm}
    \caption{\label{fig:schemes} Inverted mass hierarchy}   
\label{scheme2}
\vspace*{-2mm}
\end{figure}

For matter of constant density, closed-form solutions of the 
evolution equation can be found 
\cite{exact}; however, the corresponding expressions are rather complicated 
and not easily tractable. For a general electron density profile $N_e \ne
const$ no closed-form solutions exist. It is therefore desirable to 
have approximate analytic solutions of the neutrino evolution equation.
A number of such solutions were found, most of them based on the expansions 
in one (or both) of the two small parameters: 
\be
\Delta m_{21}^2/\Delta m_{31}^2 =
\Delta m_{\odot}^2/\Delta m_{\rm atm}^2
\aprle 0.1 \,,
\ee
\be
|U_{e3}| = |\sin\theta_{13}| \aprle 0.2 ~~\cite{Chooz} \,.
\label{small}
\ee
Our numbering of neutrino mass eigenstates corresponds to that in Figs. 
\ref{scheme1} and \ref{scheme2}, which also show schematically the
possible neutrino mass hierarchies and the flavour composition of neutrino 
mass eigenstates. 

In the limits $\Delta m_{21}^2=0$ or $U_{13}=0$ $~$ the transition 
probabilities acquire an effective 2f form. 
When both these parameters vanish, the genuine 2f case is recovered. 

\subsection{Constant-density matter}
In the case of matter of constant density approximate solutions of the
neutrino evolution \mbox{equation} were found using the expansion in
$\alpha\equiv \Delta m_{\odot}^2/\Delta m_{\rm atm}^2$ in \cite{appr1}. An 
expansion in 
both $\alpha$ and $\sin\theta_{13}$ was used in \cite{appr2}. The 
$\nu_e\leftrightarrow \nu_\mu$ transition probability found in
\cite{appr2} has the general form 
\bea
P(\nu_e\leftrightarrow \nu_\mu)\sim s_{23}^2 \tilde{P}_2(\Delta m _{31}^2, 
\theta_{13},N_e) ~+~~~~~~~~
\nonumber \\
c_{23}^2 \tilde{P}_2(\Delta m _{21}^2, \theta_{12}, N_e) 
~+~\mbox{interf. term}\,,
\label{general}
\eea
where the quantities $\tilde{P}_2$ are the 2f transition probabilities in
matter depending on the corresponding parameters shown in the parentheses. 
The interference term, which is linear in both $\alpha$ and $\sin\theta_{13}$, 
describes the genuine 3f effects, both CP-conserving and CP-violating.  

\subsection{Arbitrary density profile}
Matter of constant density is a good first approximation for long-baseline   
accelerator neutrino experiments (neutrinos traverse the mantle of the
Earth). However, it is not very useful for describing the oscillations of 
solar, atmospheric and supernova neutrinos.  
An alternative approach is to consider matter with an arbitrary density
profile and reduce the problem to an effective 2f one plus easily calculable 
3f corrections. This has been done using the expansion in $\alpha$ 
in \cite{ADLS} and the expansion in $\sin\theta_{13}$ in 
\cite{PS1999,AHLO,PS2002}. A different approach, 
based on the adiabatic approximation, was employed, e.g., in \cite{adiab}.

\begin{figure}[t] 
\vspace*{-6mm}
\epsfxsize=6.5cm\epsfbox{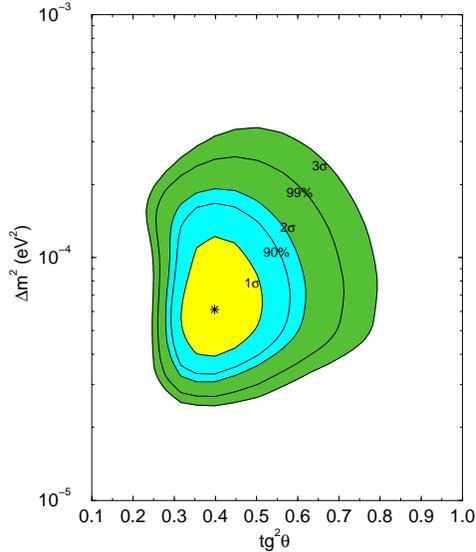}
\vspace*{-14mm}
\caption{ LMA allowed parameter region 
for $\theta_{13}=0$ \cite{dHS}  }
\label{allowed1}
\vspace*{-2mm}
\end{figure}

\subsection{Matter effects in $\nu_\mu \leftrightarrow \nu_\tau$ 
oscillations} 

Since the matter-induced potentials for $\nu_\mu$ and $\nu_\tau$ 
are the same (neglecting the radiative corrections), in the 2f case 
the $\nu_\mu \leftrightarrow \nu_\tau$ oscillations are not affected by 
matter. This, however, is not true in the 3f case; therefore matter effects 
on $\nu_\mu \leftrightarrow \nu_\tau$ oscillations is a pure 3f effect. 
It vanishes only when both $\Delta m_{21}^2$ and $U_{e3}$ vanish. 

\section{3f EFFECTS IN $\nu$ OSCILLATIONS}

We shall now discuss 3f effects in oscillations of neutrinos from various 
sources.

\subsection{Solar neutrinos }
In the 3f case, solar $\nu_e$ can in principle oscillate into either 
$\nu_\mu$, or $\nu_\tau$, or some their combination. 
What do they actually  oscillate to? 
 
It is easy to answer this question. 
The smallness of the mixing parameter $|U_{e3}|$ implies that the  
mass eigenstate $\nu_3$ 
is approximately given by 

\begin{figure}[t] 
\vspace*{-6mm}
\epsfxsize=6.5cm\epsfbox{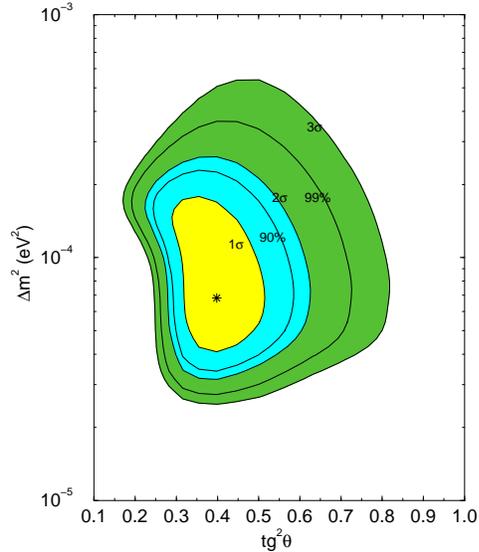}
\vspace*{-14mm}
\caption{ LMA allowed parameter region 
for $\sin^2\theta_{13}=0.04$ \cite{dHS}  }
\label{allowed2}
\vspace*{-2mm}
\end{figure}
%
\be
\nu_3 \simeq s_{23}\, \nu_\mu + c_{23}\,\nu_\tau 
\label{nu3}
\ee
and, to first approximation, does not participate in the solar neutrino 
oscillations. From the unitarity of the leptonic mixing matrix it then
follows that the solar neutrino oscillations are the oscillations between 
$\nu_e$ and a state $\nu'$ which is the linear combination of $\nu_\mu$ and 
$\nu_\tau$, orthogonal to $\nu_3$: 
\be
\nu' = c_{23}\,\nu_\mu - s_{23}\,\nu_\tau 
\label{nuprime}
\ee
Since the mixing angle $\theta_{23}$, responsible for the atmospheric 
neutrino oscillations, is known to be close to $45^\circ$, Eq. (\ref{nuprime}) 
implies that the solar $\nu_e$ oscillate into a superposition of $\nu_\mu$
and $\nu_\tau$ with equal or almost equal weights. 

What are the 3f effects in the oscillation probabilities?  Since at low 
energies $\nu_\mu$ and $\nu_\tau$ are experimentally indistinguishable, all 
the observables depend on just one probability -- the $\nu_e $ survival 
probability $P(\nu_e\to \nu_e)$.  
Averaging over fast oscillations due to the large mass squared difference  
$\Delta m_{\rm atm}^2=\Delta m_{31}^2$ yields \cite{Lim}
\be
P(\nu_e\to \nu_e) \simeq c_{13}^4 \tilde{P}_{2ee}(\Delta m_{21}^2, 
\theta_{12}, N_{\rm eff}) + s_{13}^4\,.
\label{Psol}
\ee
Here $\tilde{P}_{2ee}(\Delta m_{21}^2,\theta_{12}, N_{\rm eff})$ is the 2f
survival probability of $\nu_e$ in matter with the effective 
electron density $N_{\rm eff}=c_{13}^2\,N_e$. 

As follows from the CHOOZ data \cite{Chooz}, the second term 
in Eq. (\ref{Psol}), $s_{13}^4$,  does not exceed $10^{-3}$, i.e. is 
negligible. 
At the same time, the coefficient $c_{13}^4$ of $\tilde{P}_{2ee}$ in the 
first term may differ from unity by as much as $\sim 5$ -- 10\,\%. Thus, 
3f \mbox{effects} may lead to an energy-independent suppression of the 
$\nu_e$ survival probability by up to 10\%.  With high precision solar data 
this must be taken into account.  This is illustrated by Figs. 
\ref{allowed1} and \ref{allowed2} \cite{dHS}: The difference between 
the cases $\theta_{13}=0$ and $\sin^2\theta_{13}=0.04$ (which is 
about the maximum allowed by CHOOZ value) is quite noticeable. 

\subsection{Atmospheric neutrinos }

(1) The dominant channel $\nu_\mu \leftrightarrow \nu_\tau$. 
In the 2f limit, there are no matter effects in this channel (neglecting 
tiny $V_{\mu\tau}$ caused by radiative corrections). The oscillation 
probability is independent from the sign of $\Delta m_{31}^2$, 
i.e. cannot differentiate between the normal and inverted 
neutrino mass hierarchies. The 3f effects result in a weak sensitivity 
to matter effects and to the sign of $\Delta m_{31}^2$.  

(2) The subdominant channels $\nu_e \leftrightarrow \nu_{\mu,\tau}$. 
Contributions of these oscillation channels to the number of $\mu$ -- 
like events are subleading and \mbox{difficult} to observe.  For e-like
events, one could {\it a priori} expect significant oscillations effects.  
However, these effects are in fact strongly suppressed because of the 
specific composition of the atmospheric neutrino flux and 
proximity of the 
mixing angle $\theta_{23}$ to $45^\circ$.  
Indeed, in the 2f limits one finds 
\be
~~\frac{F_e-F_e^0}{F_e^0} = \tilde{P}_2(\Delta m_{31}^2,\,\theta_{13},
V)\cdot (r s_{23}^2-1)
\label{relat1}
\ee
in the limit $\Delta m_{21}^2\to 0$ \cite{ADLS}, and  
\be
~~\frac{F_e-F_e^0}{F_e^0} = \tilde{P}_2(\Delta m_{21}^2,\,\theta_{12},
V)\cdot (r c_{23}^2-1)
\label{relat2}
\ee
in the limit $s_{13}\to 0$ \cite{PS1999}. 
Here $F_e^0$ and $F_e$ are the $\nu_e$ fluxes in the absence and in the 
presence of the oscillations, respectively, and $r \equiv F_\mu^0/F_e^0$.  
At low energies $r \simeq 2$; also, we know that $s_{23}^2\simeq c_{23}^2 
\simeq 1/2$. Therefore the factors $(r s_{23}^2-1)$ and $(r c_{23}^2-1)$ 
in Eqs. (\ref{relat1}) and (\ref{relat2}) are very small and strongly
suppress the oscillation effects even if the transition probabilities 
$\tilde{P}_2$ are close to unity. This happens because of the strong 
cancellations of the transitions from and to the $\nu_e$ state. 

All this looks as a conspiracy to hide the oscillation effects on the 
e-like events! This conspiracy is, however, broken by the 3f effects. 
Keeping both $\Delta m_{21}^2$ 
and $s_{13}$ in leading order yields \cite{PS2002}
\bea
\frac{F_e-F_e^0}{F_e^0} &\simeq& \tilde{P}_2(\Delta 
m_{31}^2,\,\theta_{13}) \cdot (r\,s_{23}^2-1) \\ 
&+& \tilde{P}_2(\Delta m_{21}^2,\,\theta_{12}) \cdot  (r\,c_{23}^2-1) 
\label{relat3}
\eea
\[
\quad\quad\quad\quad  \quad -~~2 s_{13}\, s_{23}\, c_{23}\, r\, {\rm Re}
(\tilde{A}_{ee}^* \,
\tilde{A}_{\mu e})
\]
\vspace{3mm}
The interference term, which represents the genuinely 3f effects, is not 
suppressed by the flavour composition of the atmospheric neutrino flux; it
may be responsible (at least, partially) for some excess of the upward-going 
sub-GeV e-like events observed at Super-Kamiokande \cite{PS2002}.  

\subsection{Reactor antineutrinos}
Since the average energy of reactor $\bar{\nu}_e$'s is $\bar{E} \sim 3$ 
MeV, for intermediate-baseline experiments, such as CHOOZ and Palo 
Verde ($L\sim 1$ km), one has 
\be 
\frac{\Delta m_{31}^2}{4E}\,L\sim 1\,, \quad\quad 
\frac{\Delta m_{21}^2}{4E}\,L \ll 1 \,.
\label{cond1}
\ee
This justifies the use of the one mass scale dominance approximation, 
which gives 
\be
P(\bar{\nu}_e\to\bar{\nu}_e) = 1- \sin^2 2\theta_{13} \cdot
\sin^2\left(\!\frac{\Delta m_{31}^2}{4E}L\!\right),
\label{Preac1}
\ee
a pure 2f result. However, in the case of the LMA solution of the solar 
neutrino problem, at high \mbox{enough} confidence level $\Delta m_{21}^2$ 
can be comparable with $\Delta m_{31}^2$, and the second condition in (12) 
may not be valid. In such a situation the 3f effects coming through the 
subdominant $\Delta m_{21}^2$ should be taken into account, The analyses 
\cite{highLMA} show that the constraints on $|U_{e3}|$ derived from  
the CHOOZ experiment become slightly more stringent in that case. 
However, the new SNO data \cite{newSNO} disfavour large values of 
$\Delta m_{21}^2$ and so make this possibility less likely. 

For KamLAND, which is a very long baseline reactor experiment 
($\bar{L}\simeq 170$ km), one has 
\be
\frac{\Delta m_{31}^2}{4E}\,L\gg 1\,, \quad
\frac{\Delta m_{21}^2}{4E}\,L \aprge 1 \quad \mbox{(for LMA)}\,. 
\label{cond2}
\ee
Averaging over the fast oscillations driven by 
$\Delta m_{31}^2=\Delta m_{\rm atm}^2 $ yields  
\be
P(\bar{\nu}_e\to \bar{\nu}_e) = c_{13}^4 P_{2 \bar{e} \bar{e}}(\Delta
m_{21}^2, \theta_{12}) + s_{13}^4\,.
\label{Preac2}
\ee
This has the same form as Eq. (\ref{Psol}), except that the 2f survival
probability $P_{2 \bar{e} \bar{e}}$ has to be calculated in vacuum rather 
than in matter; it is in fact given by Eq. (\ref{Preac1}). The probability
(\ref{Preac2}) can differ from the 2f probability (\ref{Preac1}) by 
up to $\sim 10\%$. 

\subsection{LBL accelerator experiments}

(1) $\nu_\mu$ disappearance. 

3f effects can result in up to $\sim 10$\% corrections to the disappearance 
probability, mainly due to the factor $c_{13}^4$ in the effective 
amplitude of the $\nu_\mu\leftrightarrow \nu_\tau$ oscillations, 
\be
\sin^2 (2\theta_{\mu\tau})_{\rm eff} = c_{13}^4
\,\sin^2 2\theta_{23}\,.
\label{amplit}
\ee
Another manifestation of 3-flavourness are small matter effects in 
$\nu_\mu\leftrightarrow \nu_\tau$ oscillations. The same applies to 
$\nu_\tau$ appearance in experiments with the conventional neutrino
beams.  
$\nu_\mu$ disappearance also receives contributions from the subdominant 
$\nu_\mu\leftrightarrow \nu_e$ oscillations. 

(2) $\nu_\mu$ appearance at neutrino factories; $\nu_e$ appearance at 
neutrino factories and in experiments with the conventional neutrino beams. 

These are driven by the $\nu_e \leftrightarrow \nu_{\mu,\tau}$ 
\mbox{oscillations}. 
There are two channels through which these subdominant oscillations can 
proceed -- those governed by the parameters $(\theta_{13},~\Delta m_{31}^2)$ 
and $(\theta_{12},~\Delta m_{21}^2)$. For typical energies of the LBL 
accelerator experiments (a few GeV to tens of GeV), and assuming the LMA 
solution of the solar neutrino problem, one finds that for $\theta_{13}$ 
in the range $3\cdot 10^{-3}\,\aprle\, \theta_{13}\, \aprle\, 3\cdot 10^{-2}$ 
the two channels compete; otherwise one of them dominates.  

Unlike in the case of atmospheric neutrinos, there is no suppression 
of the oscillation effects on the $\nu_e$ flux due to the flavour composition 
of the original flux.

The dependence of the oscillation probabilities on the CP-violating phase 
$\delta_{\rm CP}$ (both $\sim \sin\delta_{\rm CP}$ and $\sim 
\cos\delta_{\rm CP}$) comes from the interference terms and is a pure 3f 
effect. 
The 3f effects will be especially important for the future experiments
at neutrino factories which are designed 
for precision measurements of neutrino parameters.

\subsection{Supernova neutrinos}

\begin{figure}[hbt]  
\psfrag{r}{{\small $n_e$}}
\psfrag{a}{{\small $\nu_e$}}   
\psfrag{b}{{\small $\nu_{\tau}$}}  
\psfrag{c}{{\small $\nu_{\mu}$}}
\psfrag{d}{{\small $\nu_{3 m}\quad $}}
\psfrag{e}{{\small $\nu_{2 m}\quad $}}
\psfrag{f}{{\small $\nu_{1 m}$}}
\psfrag{g}{{\small $\nu_{\mu}'$}}
\psfrag{h}{{\small $\nu_{\tau}'$}}
\psfrag{j}{{\small $\bar{\nu}_{\mu}$}} 
\psfrag{k}{{\small $\bar{\nu}_{\tau}$}}
\psfrag{w}{{\small $\bar{\nu}_e$}}
\psfrag{X}{{\small $\mu \tau$}}
\epsfig{file=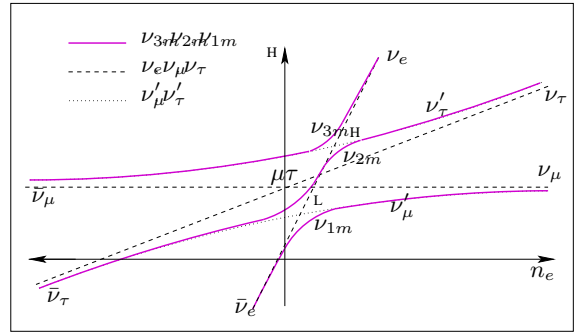, width=7.5truecm}
\vspace*{-8mm}
\caption{Energy level crossing scheme for supernova neutrinos}
\label{levcross}
\end{figure}

In supernovae, matter density varies in a very wide range, and the conditions 
for the three \mbox{MSW} resonances are satisfied (taking into account that 
due to radiative corrections $V_{\mu\tau}\ne 0$), see Fig.~\ref{levcross}. 
The hierarchy $\Delta m^2_{21}\ll \Delta m^2_{31}$ leads to the approximate 
factorization of transition dynamics at the resonances, so that the 
transitions, to first approximation, are effectively 2f ones. However, the 
observable effects of the supernova neutrino oscillations depend 
on the transitions between all three neutrino species. 

The Earth matter effects on supernova neutrinos can be used to measure
$|U_{e3}|$ to a very high accuracy ($\sim 10^{-3}$) and to determine the 
sign of $\Delta m_{31}^2$  \cite{SN1}. 

The transitions due to the $\nu_\mu - \nu_\tau$ potential 
\mbox{difference} $V_{\mu\tau}$ caused by radiative corrections may 
have observable consequences if the originally produced $\nu_\mu$ and
$\nu_\tau$ fluxes are not exactly the same \cite{SN2}.
%
\section{CP AND T VIOLATION IN \\ $\nu$ OSCILLATIONS IN VACUUM}

The probability of $\nu_a \to \nu_b$ oscillations in vacuum is given by
\be
P(\nu_a,t_0 \to\nu_b; t)=\left|\sum_i U_{bi} e^{-i E_i (t-t_0)}
U_{ai}^*\right|^2\!. 
\label{vac}
\ee
In the general case of $n$ flavours the leptonic mixing matrix $U_{ai}$ 
depends on $(n-1)(n-2)/2$ Dirac-type CP-violating phases $\{\delta_{\rm CP}\}$. 
If neutrinos are Majorana particles, there are $n-1$ additional, so-called 
Majorana-type CP-violating phases. However, they do not affect neutrino
oscillations and therefore I shall not discuss them. 
 
Under CP transformation, neutrinos are replaced by their antiparticles
($\nu_{a,b}\leftrightarrow \bar{\nu}_{a,b}$), which is equivalent to 
the complex conjugation of $U_{ai}$:
\bea
{\rm CP:}\quad \nu_{a,b}\leftrightarrow \bar{\nu}_{a,b}\qquad\qquad
\qquad\qquad\qquad\qquad
\nonumber \\
\Leftrightarrow~ U_{ai} \to U_{ai}^* \quad(\{\delta_{\rm CP}\} \to
-\{\delta_{\rm CP}\})\,. ~~~
\label{CPvac}
\eea

Time reversal transformation interchanges the initial and final evolution 
times $t_0$ and $t$ in \mbox{Eq.} (\ref{vac}),  i.e. corresponds to evolution 
``backwards in time''. As follows from Eq. (\ref{vac}), the interchange 
$t_0 \buildrel \rightarrow \over {_{\leftarrow}} t$ is equivalent to the 
complex conjugation of the exponential factors in the oscillation 
amplitude. Since the transition probability only depends on the modulus of 
the amplitude, this is equivalent to the complex conjugation of the factors 
$U_{bi}$ and $U_{ai}^*$, which in turn amounts to interchanging 
$a\buildrel \rightarrow \over {_{\leftarrow}} b$. Thus, instead of
evolution ``backwards in time'' one can consider evolution forward in
time, but between the interchanged initial and final flavours: 
\bea
T: ~~t_0\buildrel \rightarrow \over {_{\leftarrow}} t 
\Leftrightarrow \nu_{a} \leftrightarrow \nu_{b}  
\qquad\qquad\qquad\qquad\qquad
\nonumber \\
\Rightarrow  ~U_{ai} \to U_{ai}^*
~~(\{\delta_{\rm CP}\} \to - \{\delta_{\rm CP}\})\,. 
\qquad
\label{Tvac}
\eea

Under the combined action of CP and T: 
\bea
{\rm CPT}: \quad \nu_{a,b} \leftrightarrow \bar{\nu}_{a,b} ~~\&~~ 
t_0\buildrel \rightarrow \over {_{\leftarrow}} t 
~(\nu_a \leftrightarrow \nu_b) 
\nonumber \\ 
\quad\quad\quad\quad\quad \Rightarrow ~P(\nu_a\to\nu_b) \to P(\bar{\nu}_b
\to \bar{\nu}_a)\,.~
\label{CPT}
\eea
{}From CPT invariance it follows that CP violation
implies T violation and vice versa. 

CP and T violation can be characterized by the probability differences 
\be
\Delta P_{ab}^{\rm CP}\equiv P(\nu_a\to \nu_b)-P(\bar{\nu}_a\to\bar{\nu}_b)\,,
\label{DCP}
\ee
\be
\Delta P_{ab}^{\rm T}\, \equiv\, P(\nu_a\to \nu_b)-P(\nu_b\to\nu_a)\,.
\label{DT}
\ee
{}From CPT invariance it follows that the CP- and T-violating probability 
differences coincide, and that the survival probabilities 
have no CP asymmetry: 
\be
\Delta P_{ab}^{\rm CP} = \Delta P_{ab}^{\rm T}\,; 
\quad\quad \Delta P_{aa}^{\rm CP} = 0\,.
\label{CPT2}
\ee
CP and T violations are absent in the 2f case, so any observable violation
of these symmetries in neutrino oscillations in vacuum would be a pure
$\ge 3$f effect. 

In the 3f case, there is only one CP-violating Dirac-type phase
$\delta_{\rm CP}$ and so only one CP-odd (and T-odd) probability
difference: 
\be
\Delta P_{e\mu}^{\rm CP} ~=~ \Delta P_{\mu\tau}^{\rm CP} ~=~ 
\Delta P_{\tau e}^{\rm CP} ~\equiv~ \Delta P\,,
\label{DeltaP}
\ee
\[
\Delta P ~=~ {} -\, 4 s_{12}\,c_{12}\,s_{13}\,c_{13}^2\,s_{23}\,c_{23}\,
\sin\delta_{\rm CP}\,\times 
\quad \quad \quad \quad \quad \quad \quad \quad \quad
\]
\[
\left[\sin\left(\!\frac{\Delta m_{12}^2}{2E} L\!
\right)\!+\sin\left(\!\frac{\Delta m_{23}^2}{2E} L\!\right) \!+ \sin\left(
\!\frac{\Delta m_{31}^2}{2E} L\!\right)\! \right].
\]
It vanishes 

$\bullet$ when at least one $\Delta m_{ij}^2 = 0$

$\bullet$ when at least one $\theta_{ij} = 0$ or $90^\circ$ 

$\bullet$ when $\delta_{\rm CP} = 0$ or $180^\circ$

$\bullet$ in the averaging regime 

$\bullet$ in the limit $L\to 0$ (as $L^3$) \\
Clearly, this quantity is very difficult to observe.  

\section{CP AND T VIOLATIONS IN \\ $\nu$ OSCILLATIONS IN MATTER}

For neutrino oscillations in matter, CP transformation (substitution 
$\nu_a\leftrightarrow \bar{\nu}_a$) implies not only complex conjugating 
the leptonic mixing matrix, but also flipping the sign of the 
matter-induced neutrino potentials: 
\bea
{\rm CP:} \quad 
U_{ai} \to U_{ai}^* ~(\{\delta_{\rm CP}\} \to - \{\delta_{\rm CP}\})\,,
\nonumber \\
V(r) \to {}- V(r)\,.
\qquad\quad~  
\label{CPmat}
\eea

It can be shown \cite{AHLO} that in matter with an arbitrary density profile, 
as well as in vacuum, the action of time reversal on neutrino oscillations
is equivalent to interchanging the initial and final neutrino flavours. It 
is also equivalent to complex conjugating $U_{ai}$ and replacing the matter 
density profile by the reverse one: 
\bea
{\rm T:} \quad
%
U_{ai} \to U_{ai}^* ~(\{\delta_{\rm CP}\} \to - \{\delta_{\rm CP}\})\,,
\nonumber \\
V(r) \to \tilde{V}(r)\,.  ~~\quad\quad\quad\quad
\label{Tmat}
\eea
Here 
\be
\tilde{V}(r) = \sqrt{2} G_F \tilde{N} (r)\,,
\label{Vrev}
\ee
$\tilde{N} (r)$ being the reverse profile, i.e. the profile that corresponds 
to the interchanged positions of the neutrino source and detector. In the 
case of symmetric matter density profiles (e.g., matter of constant density), 
$\tilde{N} (r) = N (r)$. 

An important point is that the very presence of matter (with unequal numbers
of particles and antiparticles) violates C, CP and CPT, leading to CP 
violation in neutrino oscillations even in the absence of the fundamental 
CP-violating phases $\{\delta_{\rm CP}\}$. This fake (extrinsic) CP
violation 
may complicate the study of the fundamental (intrinsic) one.

\subsection{CP violation in matter}

Unlike in vacuum, CP violation in neutrino oscillations in matter exists 
even in the 2f case (in the case of three or more flavours, even when all
$\{\delta_{\rm CP}\}=0$):  
\be 
P(\nu_a\to \nu_b) \ne P(\bar{\nu}_a\to \bar{\nu}_b) \,.
\label{diff}
\ee
This is actually a well known fact -- for example, the MSW effect can
enhance the $\nu_e\leftrightarrow \nu_\mu$ oscillations and suppress the 
$\bar{\nu}_e\leftrightarrow \bar{\nu}_\mu$ ones or vice versa.
Moreover, in matter the survival probabilities are not CP-invariant:
\be
P(\nu_a\to \nu_a) \ne P(\bar{\nu}_a\to \bar{\nu}_a) \,.
\label{CPnoninv}
\ee
To disentangle fundamental CP violation from the matter induced one in
the LBL experiments one would need to measure the energy dependence of the 
oscillated signal or the signals at two baselines, which is a difficult 
task. The (difficult) alternatives are: 

$\bullet$ 
LBL experiments at relatively low energies and moderate baselines ($E\sim$
0.1 -- 1 GeV, $L\sim$ 100 -- 1000 km) \cite{lowE}. 

$\bullet$ Indirect measurements through \\ 
$~~~~~~\,$ (A) CP-even terms $\sim \cos\delta_{\rm CP}$ \cite{CPeven};  

\noindent
$~~~~\;$ (B) Area of leptonic unitarity triangle \cite{triangle}. 
CP violation cannot be studied in the supernova neutrino experiments
because of the experimental indistinguishability of low-energy $\nu_\mu$
and $\nu_\tau$.

\subsection{T violation in matter}

Since CPT is not conserved in matter,  CP and T violations are no longer 
directly connected (although some relations between them still exist 
\cite{AHLO,MNP}). 
Therefore T violation in neutrino oscillation in matter deserves 
an independent study. Its characteristic features are:  

$\bullet$ Matter does not necessarily induce T violation (only asymmetric 
matter with $\tilde{N}(r) \ne N(r)$ does).  

$\bullet$ There is no T violation (either fundamental or matter 
induced) in the 2f case. This is a simple consequence of unitarity.
For example, for the $(\nu_e, \nu_\mu)$ system one has 
\bea 
P_{ee} + P_{e\mu} =1\,, \nonumber \\
P_{ee} + P_{\mu e} =1\,,
\label{unit1}
\eea
from which $P_{e\mu} = P_{\mu e}$. 

$\bullet$ In the 3f case there is only one T-odd probability difference
for $\nu$'s (and one for $\bar{\nu}$'s), irrespective of the matter
density profile:  
\be 
\Delta P_{e\mu}^T=\Delta P_{\mu\tau}^T=\Delta P_{\tau e}^T \,.
\label{unit2}
\ee
This is a consequence of 3f unitarity \cite{KP1988}. 

The matter-induced T violation is an interesting, pure $\ge$3f matter
effect, absent in symmetric matter (in
particular, in constant-density matter). It does not vanish in the regime
of complete averaging of neutrino oscillations \cite{AHLO}. It may fake
the fundamental T violation and complicate its study, i.e. the extraction
of $\delta_{\rm CP}$ from the experiment. The matter-induced T violation 
vanishes when either $U_{e3}=0$ or $\Delta m_{21}^2=0$ (i.e., in the 2f 
limits) and so is doubly suppressed by both these small parameters. This 
implies that the perturbation theory can be used to obtain analytic 
expressions for the T-odd probability differences. The general structure 
of these differences is  
\be
\Delta P_{e\mu}^T = \sin\delta_{\rm CP}\cdot Y 
+\cos\delta_{\rm CP} \cdot X\,.
\label{struct}
\ee
Here the first term ($\propto \sin\delta_{\rm CP}$) is due to the 
fundamental T violation, whereas the second term is due to the
matter-induced one. In the adiabatic approximation one finds \cite{AHLO} 
$X = J_{\rm eff}\times\mbox{(oscillating terms)}\,,$ where 
\be
J_{\rm eff}~=~ {} s_{12}\,c_{12}\,s_{13}\,c_{13}^2\,s_{23}\,c_{23}\,
\frac{\sin(2\theta_1-2\theta_2)}{\sin 2\theta_{12}}\,.
\label{Jeff}
\ee
Here $\theta_{1}$ and $\theta_{2}$ are the mixing angles in matter in the
(1-2) sector at the initial and final points of neutrino evolution, 
respectively; $\theta_{1}-\theta_{2}$ is therefore a measure of the 
asymmetry of the density profile. $J_{\rm eff}$ has to be compared with the 
vacuum Jarlskog invariant 
\be
J~=~s_{12}\,c_{12}\,s_{13}\,c_{13}^2\,s_{23}\,c_{23}
\sin\delta_{\rm CP} \,.
\label{J}
\ee
We see that the factor $\sin(2\theta_1-2\theta_2)/\sin 2\theta_{12}$
in $J_{\rm eff}$ plays the same role as the factor 
$\sin\delta_{\rm CP}$ in $J$. 

In an asymmetric matter, both fundamental and matter-induced T violations 
contribute to the T-odd probability differences $\Delta P_{ab}^T$. This 
may hinder the experimental determination of the fundamental CP- and
T-violating phase $\delta_{\rm CP}$. In particular, in the accelerator 
LBL experiments one has to take into account that the Earth's density 
profile is not perfectly spherically symmetric. To extract the fundamental 
T violation, strictly speaking one would need to measure 
\be
P_{\rm dir}(\nu_a\to \nu_b) -
P_{\rm rev}(\nu_b\to \nu_a)\,, 
\label{deltaP}
\ee
where $P_{\rm dir}$ and $P_{\rm rev}$ correspond to the direct and
reverse matter density profiles. 
(An interesting point is that even the survival probabilities $P_{\mu\mu}$ 
and $P_{\tau\tau}$ can be used for that \cite{FK}. 
The $\nu_e$ survival probability $P_{ee}$ is an exception because in the
3f case it does not depend on $\delta_{\rm CP}$ \cite{KuPa,MW}. This, 
however, is not true if $\nu_{\rm s}$ is present \cite{AHLO}). 

In practical terms, it would certainly be difficult to measure the 
quantity in (\ref{deltaP}): It would not be easy, for example, to move 
CERN to Gran Sasso and the Gran Sasso Laboratory to CERN. Fortunately, this 
is not actually necessary -- matter-induced T
violation due to imperfect sphericity of the Earth's density distribution 
is very small. It cannot spoil the
determination  of $\delta_{\rm CP}$ if the error in $\delta_{\rm CP}$ is
$>1$\% at 99\% C.L. \cite{AHLO}.  

Can we study 
T violation in neutrino oscillations experimentally? Because of problems 
with the detection of $e^\pm$ this seems to be difficult, but probably
not impossible. To study matter-induced T violation would be a harder task.   
\mbox{T-odd} matter effects are expected to 
be negligible in terrestrial experiments. They cannot be observed in the 
supernova neutrino oscillations because of the experimental 
indistinguishability of low -- energy $\nu_\mu$ and $\nu_\tau$. It could, 
however, affect the signal from $\sim $GeV neutrinos produced in the 
annihilations of WIMPs inside the Sun \cite{deGouv}.

\section{A HYMN TO $U_{e3}$}

The leptonic mixing parameter $U_{e3}$ plays a very special role in
neutrino physics.  It is of particular interest for a number of reasons. 

First, it is the least known of leptonic mixing parameters: while we
have (relatively small) allowed ranges for the other two mixing parameters, 
we only know an upper bound on $|U_{e3}|$. 
\mbox{Its} smallness, which looks strange in the light of the fact that 
the other two mixing \mbox{parameters}, $\theta_{12}$ and $\theta_{23}$, are 
apparently large, remains essentially unexplained. (There are, however, some 
ideas which relate the smallness of $|U_{e3}|$ to that of $\Delta m_\odot^2/
\Delta m_{\rm atm}^2$ \cite{ABR,King}). 

The smallness of $U_{e3}$ is likely to be the bottleneck for studying
the fundamental CP and T violation effects and matter-induced
T violation in neutrino oscillations. 
%
The same applies to the determination of the sign of $\Delta m_{31}^2$ in 
future LBL experiments, which would allow us to discriminate between 
the normal and inverted neutrino mass hierarchies. Therefore it would be
vitally important to know how small $U_{e3}$ actually is.  

The parameter $U_{e3}$ can be efficiently used to discriminate between 
various neutrino mass models \cite{Barr,Tan}. 
It is one of the main parameters that 
%
drives the subdominant oscillations of atmospheric neutrinos and is
important for their study. 
%
It also governs the Earth matter effects on supernova neutrino 
oscillations. 

And finally, $U_{e3}$ apparently provides us with the only opportunity to see 
the ``canonical'' \mbox{MSW} effect. While matter effects can be important 
even in the case of large vacuum mixing angles, the most spectacular
phenomenon, strong enhancement of mixing by matter, can only occur  
if the vacuum mixing angle is small. From what we know now, 
it seems that the only small leptonic mixing parameter is $U_{e3}$. 

All this makes measuring $U_{e3}$ one of the most important problems in 
neutrino physics.

\section{4f OSCILLATIONS}

If the LSND experiment is correct, the oscillations interpretation of
the solar, atmospheric and accelerator neutrino data would require
three distinct values of $\Delta m^2$: $\Delta m_\odot^2\ll \Delta m_{\rm
atm}^2\ll \Delta m_{\rm LSND}^2$. This would imply the existence of at 
least four light neutrino species, $\nu_e$, $\nu_{\mu}$, $\nu_{\tau}$ and 
$\nu_s$. A possible alternative is a strong CPT violation in the neutrino 
sector, leading to 
inequalities of $\Delta m^2$ in the neutrino and antineutrino sectors 
\cite{CPTV}; I will not discuss this possibility here. 

In general, the 4f neutrino oscillations are described by 6 mixing angles
$\theta_{ij}$, 3 Dirac-type CP-violating phases and 3 values of 
$\Delta m_{ij}^2$, i.e. are quite complicated. 
%
Fortunately, there is a simplification: The data admit only 2 classes of
4f schemes, the so-called  (3+1) and  (2+2) schemes.
%
In the (3+1) schemes, three neutrino mass eigenstates are close to each
other while the fourth one is separated from them by a large mass gap. 
This mass eigenstate is predominantly $\nu_s$ with small admixtures of the 
active neutrinos:
\be
\nu_4 \simeq \nu_s +  {\cal O}(\epsilon)\cdot 
(\nu_e, ~\nu_\mu, ~\nu_\tau) \,,\quad \epsilon \ll 1\,,
\label{nu4}
\ee
whereas $\nu_1$, $\nu_2$ and $\nu_3$ are the usual linear combinations of
$\nu_e$, $\nu_\mu$ and $\nu_\tau$ plus small ($\sim\epsilon$) admixtures of  
$\nu_s$. In this scheme the amplitude of the $\nu_\mu\rightarrow \nu_e$
oscillations at LSND is 
\be
\sin^2 2\theta_{\rm LSND} = 4\,|U_{e4}\, U_{\mu 4}|^2 \sim \epsilon^4\,.
\label{LSND1}
\ee 
Strong upper bounds on $|U_{e4}|$ and $|U_{\mu 4}|$ from $\bar{\nu}_e$ and
$\nu_\mu$ disappearance experiments make it rather difficult to fit
the LSND data in the (3+1) schemes \cite{MSV2001}.

In the (2+2) schemes, there are two pairs of mass eigenstates with 
relatively small mass squared differences, $\Delta m_\odot^2$ and 
$\Delta m_{\rm atm}^2$, between the states within the pairs and  
a large separation ($\Delta m_{\rm LSND}^2$) between the two pairs. 
The $\nu_\mu$ state is predominantly in the pair responsible for the 
atmospheric neutrino oscillations, whereas $\nu_e$ is mainly in the
pair responsible for $\nu_\odot$ oscillations: 
\[
\nu_{\rm atm} ~{\rm osc.:} \quad \nu_\mu\leftrightarrow \nu'\,, 
\qquad
\nu_\odot ~{\rm osc.:} \quad \nu_e \leftrightarrow \nu''\,,
\]
where  
\bea
\nu' ~\simeq~ c_\xi\,\nu_\tau + s_\xi \,\nu_s + {\cal O}(\epsilon)\cdot
\nu_e\,,
\qquad~\nonumber \\
\nu'' \simeq -s_\xi\,\nu_\tau + c_\xi\,\nu_s 
+ {\cal O}(\epsilon)\cdot \nu_\mu\,.
\qquad
\label{nuatmnusol}
\eea
The amplitude of the $\nu_\mu\rightarrow \nu_e$ oscillations at 
\mbox{LSND} is 
\be
\sin^2 2\theta_{\rm LSND} \sim \epsilon^2 \,.
\label{LSND2}
\ee 
It is only of the second order in $\epsilon$ and, unlike in the (3+1) case,
the LSND data can be easily fitted. 

However, the (2+2) schemes suffer from a different problem. The fractions of 
$\nu_s$ involved in the oscillations of atmospheric and solar 
neutrinos must sum to unity in these schemes \cite{PS2000}: 
\be
|\langle\nu_s|\nu''\rangle|^2~+~|\langle\nu_s|\nu'\rangle|^2 \simeq
c_\xi^2+s_\xi^2=1 \,.
\label{nu2primes}
\ee
This sum rule is in conflict with the atmospheric and solar neutrino data. 
Indeed, the Super-Kamiokande atmospheric neutrino data lead to the upper 
limits $\sin^2\xi<0.20$ at 90\% C.L. and $\sin^2\xi <0.26$ at 99\% C.L. 
\cite{shiozawa}. At the same time, the (pre-SNO neutral current) 
solar neutrino data data imply 
$\sin^2\xi > 0.7$ (90\% C.L.); $\sin^2\xi > 0.48$ (99\% C.L.) for the LMA  
solution of the solar neutrino problem \cite{GGMPG}. 
The recently published SNO neutral current data \cite{newSNO} will
probably strengthen this limit. Therefore, the (2+2) scenarios are also 
strongly disfavoured by the data. 

In the 4f case, there may be interesting matter effects on neutrino 
oscillations \cite{DGKK}.  CP violation is potentially much richer than
in the 3f case: there are several CP-violating observables, and large 
CP-odd effects are possible (in general, there is no suppression
due to small $\Delta m_\odot^2$). Also large T violation (both fundamental 
and matter-induced) can occur.  

\section{CONCLUSIONS}

3f effects in solar, atmospheric, reactor and supernova neutrino 
oscillations and in LBL accelerator neutrino experiments may be quite
important. They can lead to up to $\sim 10$\% corrections
to the oscillation probabilities and also to specific effects, absent 
in the 2f case. The manifestations of $\ge 3$ flavours in neutrino
oscillations include fundamental CP violation
and T violation, matter-induced T violation, matter effects in $\nu_\mu
\leftrightarrow \nu_\tau$ oscillations, and specific CP- and T-conserving
interference terms in oscillation probabilities. The leptonic mixing
parameter $U_{e3}$ plays a very special role and its study is of great 
interest. 

In the 4f case, large CP violation and (both fundamental and
matter-induced) T violation effects are possible. However, 4f scenarios 
are strongly disfavoured by the data.

\end{document}